\documentclass{llncs}
\pdfoutput=1
\usepackage{url}
\usepackage{enumerate}
\usepackage{amsmath,amssymb,amstext}

\newcommand{\NP}{\ensuremath{\mathsf{NP}}}
\renewcommand{\P}{\ensuremath{\mathsf{P}}}

\title{A Survey on Methods and Systems for Graph Compression}
\author{Sebastian Maneth \and Fabian Peternek}
\institute{University of Edinburgh\\\email{smaneth@inf.ed.ac.uk}, \email{f.peternek@ed.ac.uk}}
\date{31. March 2015}

\begin{document}
\maketitle
\begin{abstract}
    We present an informal survey (meant to accompany another paper) on graph compression methods.
    We focus on lossless methods, briefly list available approaches, and compare
    them where possible or give some indicators on their compression ratios. We also mention
    some relevant results from the field of lossy compression and algorithms specialized for the use
    on large graphs. --- \textbf{Note:} The comparison is by no means complete. This document is a
    first draft and will be updated and extended.
\end{abstract}
\section{Introduction}
Graphs are an important data structure in computer science. A graph consists of a set of nodes
$V$ and a set of edges $E$ connecting the nodes. We say that two nodes are \emph{adjacent}, if they
are connected by an edge. The nodes and edges may be labeled. The edges may be directed or
undirected. Furthermore sometimes multigraphs are considered, which allow for more
than one edge between two nodes (called multiedges). Here we want to consider how graphs are
represented in memory and how they are stored in external memory. This is of renewed interest,
as many recent graph data is too large to fit into memory. Compression can be applied to fit large
graphs in memory and to store them efficiently on secondary memory.

Let $n = |V|$ and $m = |E|$. The most basic model of a graph, is one without edge
labels and multiedges. Such graphs can be represented as:
\begin{description}
    \item[Adjacency matrix:] a square 0/1-matrix with one row/column for every node. A 1-value at
        position $i,j$ means that there is an edge from node $i$ to node $j$. Uses $O(n^2)$ bits of
        space and deciding adjacency is possible in constant time.
    \item[Adjacency list:] Represents just the edges by listing the neighbors of every nodes in a
        list. Can be imagined as an adjacency matrix that without the 0-values. Uses $O(
        (n+m)\log_2n)$ space and deciding adjacency takes $O(\log_2 n)$ time.
\end{description}
Note that both representations automatically provide node IDs, i.e. every node can be uniquely
identified by an integer assigned to it. For us, a compressed graph is a representation that uses
less space than the above two. Compression methods are often compared by the amount of bits they
need on average to store one edge (bits per edge, \emph{bpe}). To give a comparison: an adjacency
list representation of a graph with at least as many edges as nodes and node IDs represented by
32~bit integers would roughly need 32 to 64~bpe.

Most compression approaches we mention support simple queries on the compressed
graph, without the need for decompression. The most common such queries are the sets of
out-neighbors or in-neighbors of a node $v$. If there is an edge from $u$ to $v$ then $v$ is an
\emph{out-neighbor} of $u$ and $u$ is an \emph{in-neighbor} of $v$. Common traversal mechanisms like
BFS or DFS can be implemented using in- or out-neighbors. 

\emph{Note:} If not stated otherwise, the methods discussed in this survey assume directed graphs
with unique node IDs and without edge-labels or multiedges.
\section{Lossless Graph Compression}
We begin with compression methods that can be reversed. We divide them into two
categories: (1) succinct representations encode the graph in a succinct bitstring, but do not make
structural changes that have to be reversed for decompression, (2) Structural approaches replace
repeating structures by a short identifier. Note that these two approaches are often combined.
\subsection{Succinct Representation}
Succinct data structures typically mean encodings that represent the data within the information
theoretic lower bound plus $o(n)$ additional bits. We relax this definition in the second
subsection, where such proofs of optimality are not given.
\subsubsection{Without Implementation}
To our knowledge the following methods are investigated theoretically and have not been
implemented. Most of them work only on certain families of graphs, not arbitrary ones.

Galperin and Wigderson~\cite{DBLP:journals/iandc/GalperinW83} mention succinct graph representations
are mentioned by, where they examine the complexity of certain graph problems under the assumption
that the input is given in a succinct way. In this context that means, that the input size is not
polynomial in $|V|$, but polylog in $|V|$. They show that every nontrivial problem they examine
becomes $\NP$-hard in this setting, even the ones that were in \P for standard representations. How
to achieve a succinct representation is not mentioned in the paper.

Tur\'an~\cite{Turan84_planarGraphComp} presents a particular succinct representation. It encodes
undirected planar graphs without loops or multiple edges into a bitstring of length $\leq 12 n$.
Furthermore an undirected planar graph with node labels can be represented by $n\lceil \log_2
n\rceil+12n$ bits. This is achieved by first choosing a spanning tree of the graph. The nodes of
the graph are then arranged into a cyclic sequence according to a post-order traversal of the
spanning tree (length: $2n-2$). The remaining edges (i.e. the ones not part of the spanning tree)
are included as diagonals of the cycle. This is where planarity is necessary, as otherwise
these diagonals might cross. The sequence is then encoded as a string consisting of the symbols
$+, -, (, )$, which can be encoded by 2 bits each, leading to an encoding of at most $12n - 24$
bits. In the labeled case the same encoding is used, padded to the maximum amount of bits, followed
by a dictionary for the labels. Access to the compressed structure is not considered.

Jacobson's approach~\cite{Jacobson89_firstRetrievalCompression} works on graphs of bounded
pagenumber. This means that the graph can be represented by a bounded number of pages, which is a
linear drawing of a subset of nodes with the edges connecting them drawn above so that
they do not intersect. Such a page can be encoded as a string of parentheses, leading to an encoding
for the full graph by combining the encodings of the single pages. He proves it to be optimal for
trees and within a logarithmic factor of the information theoretic lower bound for graphs of bounded
pagenumber. Jacobson also considers accessing the succinct structure and shows that an algorithm can
test adjacency of two nodes in $O(\log n)$ bit inspections for graphs with one page. For graphs with
$k$ pages, Jacobson's approach uses $O(kn)$ bits and tests adjacency in $O(k\log n)$ time.

Deo and Litow~\cite{Deo98_structuralGraphCompression} show that graphs of bounded genus $g$
(which is the same as bounded pagenumber) can be represented using $O(g+n)$ bits and $O(g+n)$
time. However, finding a minimal genus embedding for a graph $G$ is \NP-complete, but
fixed-parameter tractable, therefore finding an embedding for a fixed genus $g$ is possible in \P.
They furthermore show separator theorems for graphs defined by an excluded minor. These and similar
separator techniques are later used for compression, as further mentioned below.

The approach by Farzan and Munro~\cite{DBLP:journals/tcs/FarzanM13} works for arbitrary graphs.
They first prove that it is not possible to achieve a
representation within the information theoretic lower bound up to lower order terms unless the graph is
either too sparse (i.e. $m = o(n^\delta)$ for any constant $\delta>0$) or too dense (i.e. $m =
\omega(n^{2-\delta})$ for any constant $\delta>0$). They then present a succinct encoding supporting
adjacency and neighborhood queries in constant time. The space needed for the encoding is within a
$(1+\varepsilon)$-factor of the information theoretic lower bound for any arbitrarily small constant
$\varepsilon > 0$.

Lu~\cite{Lu14_infTheoreticalOptimalCompression} gives a compression scheme for hereditary classes of
graphs. A class of graphs is hereditary, if it is closed under taking subgraphs. Let $\mathbb{G}$ be
such a class of graphs, and let $\mathsf{num}(\mathbb{G},n)$ be the amount of graphs with $n$ nodes
in $\mathbb{G}$. They prove that, if $\log\mathsf{num}(\mathbb{G},n) = O(n)$, given a graph $G \in
\mathbb{G}$ with $n$ nodes and a a genus-$o(\frac{n}{\log^2 n})$ embedding, it is possible to
represent the graph using $\beta n+o(n)$ bits, where $\beta$ is any positive constant such that
$\log\mathsf{num}(\mathbb{G},n) \leq \beta n+o(n)$. This is achieved by using graph separators as
above.
\subsubsection{With Implementation}
The representations in this section are implemented and experimentally evaluated. Most of them
focus on compression of web graphs, which are graphs where the nodes represent websites (labeled by
the URL), and a directed edge from node $v$ to node $u$ means that $v$ has a hyperlink to $u$.
Compression methods focussing on web graphs often rely on properties specific to web graphs, such as:
\begin{description}
    \item[Locality:] most links lead to pages within the same host.
    \item[Similarity:] pages on the same host often share the same links.
\end{description}
One way to work with these properties is to order the nodes lexicographically by their URLs. This leads to
an order where the source and target of the edges tend to be close to each other with respect to
that order. Once this order is computed the node labels can be replaced by IDs indicating their
position in the order. The \emph{WebGraph framework}~\cite{DBLP:conf/www/BoldiV04} by Boldi and
Vigna is based on this order. They compress the adjacency lists using several methods. Among those
methods are is the representation of adjacent nodes by the difference to the previous value, instead
of the nodes' ID, which reduces the amount of different values to be stored due to the high locality
of web graphs. This technique is further improved by considering integer intervals appearing in the
adjacency lists, which allows for even less variance. The high similarity can be utilized by using a
reference encoding. Instead of storing the (mostly) same lists many times, a reference number and
a copy instruction is stored with an adjacency list $l$, specifying which IDs from which list are
to be copied into $l$. Finally a $\zeta$-code is applied to the adjacency lists to store them in
small space. The methods employed to transform the adjacency lists are influenced by the node ordering.
In the first version a lexicographical ordering is used, but they later improve the framework
further by proposing a different ordering~\cite{Boldi09_permutingWebGraphs}. Using these
improvements they report compression for some web graphs as low as 2.6~bpe. Furthermore the
resulting structure can be used as an in-memory data structure, preserving the ability to query the
out-neighborhood of a node. However, the time this querying takes depends on the quality of the
compression. With the highest compression ratio, the answer to queries is in the realm of
milliseconds instead of microseconds. It should be noted that network graphs that are not web graphs
do not necessarily have the same locality and similarity properties and therefore may not compress
as well, using their methods.

Apostolico and Drovandi~\cite{Apostolico09_graphCompressionBFS} propose a different encoding which
has the advantage of not relying on an the availability of some natural ordering of the nodes and is
therefore more suitable to compress general graphs, though they only evaluate it on web graphs.
They use a BFS-order of the nodes and store a succinct representation of the adjacency list using an
entropy-based encoding. The adjacency list can still be queried, achieving access times which are
competitive to the ones in the WebGraph framework. At the same time they report better
compression ratios as low as 1.83~bpe for the UK2002 graph with 18 million nodes and
298 million edges, whereas the WebGraph framework achieves 2.22~bpe on the same graph. Both of these
numbers are for highest compression, disregarding access times. The best compromise of compression
and access times is 3.00~bpe with the WebGraph framework and 2.62~bpe using this method, which is
then still slightly faster than the WebGraph framework.

The above algorithms only support out-neighborhood queries. To also query for incoming neighbors,
they need to store the graph twice, once with inverted edges. Maserrat~\cite{mas12} presents
lossless and lossy compression schemes which support both, in- and out-neighborhood queries, and
also more efficient incremental updates than the above. These schemes are evaluated on social
networks as well as web graphs, achieving between 5 and 30~bpe. Social networks tend to be
more difficult to compress than web graphs, because there are no obvious properties such as
locality or similarity to exploit for a good ordering of the nodes. A specific example is a
lifejournal graph with about 4.8 million nodes and 68.5 million edges, where this algorithm achieves
13.9~bpe, while the WebGraph framework on the same graph achieves 14.3~bpe, but only supporting
out-neighbor queries.

So far every method discussed is based on adjacency lists. An adjacency matrix based is proposed by
Brisaboa et al.~\cite{Brisaboa09_k2Trees} who consider $k^2$-trees, a compact tree structure taking
advantage of large empty areas of the adjacency matrix of the graph. The general idea is, that the
adjacency matrix is first partitioned into rectangles. Rectangles that only include $0$-values are
represented by a leaf with value $0$. The other ones are recursively further partitioned, until a
tree is produced which represents the full adjacency matrix. Their encoding is a succinct
representation of this tree. With this method, they can provide full support for both, in- and
out-neighbor queries. They can show that their method achieves better compression ratios in this
setting: they achieve 4.2~bpe for the UK2002 graph, where the WebGraph framework needs more than 6
bpe to store two copies of the same graph. The access times are slightly slower however, especially
for in-neighbors.

An interesting aspect of the $k^2$-tree method is that it can not only be used to represent
adjacency matrices, but is also applicable to basically any binary relation. This aspect is used by
\'{A}lvarez et al.~\cite{Alvarez:2010:CRG:1830252.1830255} to extend the $k^2$-tree representation
to general graph databases instead of web graphs. Specifically they represent graphs, which
have uniquely labeled nodes, while the edges are labeled from an alphabet, allowing multiedges
if differently labeled. To achieve this, they represent a labeled, directed, attributed, multigraph
$G$ using a new data structure called \emph{Compact Graph Database}. Essentially, this consists of
three $k^2$-tree representations, encoding three different relations, which are:
\begin{itemize}
    \item The binary relation between nodes and their attribute values, where the nodes are
        the rows and all possible values the columns. A ``1'' signifies that the node has the
        corresponding columns attribute value.
    \item Analogously the relation between edges and their labels can be represented with a
        $k^2$-tree in this way.
    \item The relation between nodes is mostly a standard $k^2$-tree representation. The
        difference is, that they extend the standard representation to deal with multiple
        edges between nodes.
\end{itemize}
They evaluate their method on two datasets: \emph{Youtube} has 1.2 million nodes and 5.2 million edges and
is compressed to about 47\%; \emph{Wikipedia} has 11.1 million nodes and 90.1 million edges and is
compressed to about 43\% of the file size a plain representation needs. We do not know how many bpe
this corresponds to. They compare the method to a
graph database (DEX~\cite{DBLP:conf/cikm/Martinez-BazanMGNSL07}) that does not aim to achieve a
succinct representation (in fact, it needs 2 to 4 times more space than the plain file size on the
evaluated graphs), but at providing fast access times. Expectedly, the access times of DEX are 30 to
100 times faster than the ones of this method, but they conclude that ``taking into account the
difference between both methods in terms of space requirements, the difference in navigation times
is not too large''.

\subsection{Structural Approaches}
Contrary to the previous section, these approaches analyse and in some way change the graph
structure to achieve a compressed representation. These structural changes are done so that they are
reversible. This is usually achieved by using a dictionary or grammar to describe rules on how to
reverse the changes. All of the methods discussed here have been implemented. The best results are
generally achieved by combining a structural approach with one of the succinct representations above
to represent the remaining graph.

Buehrer and Chellapilla~\cite{Buehrer08_scalableWebGraphCompression} compress by searching for
bi-cliques (complete bipartite graphs) and replacing them with virtual nodes. Their method achieves
$1.95$~bpe on the UK2002 graph, a graph with 18 million nodes and 298 million edges. The method
retains querying capabilities but no experiments on access times are described. Furthermore it is not
clearly specified if this includes in-neighbors or only out-neighbors. As finding bi-cliques is a
hard problem they present several heuristics to do so.

Even better results are reported by Asano et al.~\cite{Asano08_webGraphCompression} by using a method
exploiting frequent patterns in the adjacency matrix. The examples they give for experimental
evaluation are rather small (at most 19 million edges), but on these they achieve compression ratios of
$1.7$ to $2.7$~bpe, whereas the WebGraph framework needs $2.1$ to $4.38$~bpe on the same graphs.
According to~\cite{Claude10_compactWebGraphRep} the retrieval of edges is not efficiently possible
in this method.

The RePair-compression scheme is a popular approximation of the smallest grammar generating a given
string. Claude and Navarro~\cite{Claude10_compactWebGraphRep} apply it to graphs by compressing a
string representation of the adjacency list. Their input file format is a binary representation of
the adjacency list. The resulting grammar uses the same format. The method allows for partial
decompression and thus the answering of out-neighborhood queries. They obtain 4.23~bpe on the UK2002
graph, which is slightly worse than WebGraph, but their access times are better.
%
%
Claude and Ladra~\cite{Claude11_practicalGraphRepresentations} then combine the previous method and
the $k^2$-tree representation. They achieve a compression of 2.27~bpe on the UK2002-graph, while
retaining the ability to query for both, in- and out-neighbors.

Grabowski and Bieniecki~\cite{DBLP:conf/icmmi/GrabowskiB11} propose a merging contiguous
blocks of adjacency lists into a single ordered list. Again on the UK2002 graph they reach a
compression ratio of $1.92$~bpe. As usual this is sacrificing access time, the degree to which this
is happening can be given as a parameter. A good compromise for the UK2002 graph is $2.73$~bpe,
which is still better than the WebGraph frameworks compression, but provides roughly the same access
time. Only out-neighbors are supported.

As far as we know, the current state of the art for methods supporting in- and out-neighbor queries
is by Hern\'andez and Navarro~\cite{DBLP:journals/kais/HernandezN14} who improve the method by
Buehrer and Chellapilla by expanding it to a general search for dense subgraphs instead of just
bicliques. They then combine this with the BFS-representation by Apostolico and Drovandi and with
the $k^2$-tree representation, where the $k^2$-representation works best regarding compression, but
makes the access times slower. This achieves representations of 0.9 to 1.5~bpe on web graphs
(UK2002: 1.53~bpe). When considering queries for both incoming and outgoing edges this is the
current state of the art regarding compression. Access is possible, but comparably slow: queries
take approximately 2 to 3 times longer than with the $k^2$-tree method. If only out-neighbor queries
are considered the best space/time trade-off is not at the best compression. In this case they
compress the UK2002 graph to about 2~bpe achieving comparable access times to the method by
Grabowski and Bienicki above. However, it achieves about 1~bpe better compression when considering
bidirectional queries at comparable access times and they improve social network compression by
up to 10~bpe. 

Navlakha et al.~\cite{DBLP:conf/sigmod/NavlakhaRS08} consider graph summarization with bounded error
using the minimum description length principle. A graph summary is a graph consisting of
``supernodes'', a mapping that maps the supernodes back to actual nodes and a list of corrections.
The idea is that if there is an edge from one supernode to another, then there is an edge from every
actual node the first supernode represents to every actual node the second represents. The list of
corrections is then used to remove surplus edges this creates, or to add edges that are missing from
this method. The user can further specify an error bound, to achieve lossy compression. If this is
done, the decompressed graph can have additional edges, or missing edges compared to the original,
up to a fraction given by the bound. They define the cost of
their representation as the sum of the number of edges in the summary graph and the number of
corrections. The mapping of the supernodes are ignored, because they will generally be small.
Several algorithms to compute such a representation are proposed, of which a greedy method achieves
the best result. Comparing the cost percentage (how much lower the cost of the representation is to
the original) they achieve compression of 30\% to 80\%, consistently better by about 10\% to 20\%
than WebGraph. Access to the compressed structure is not considered. Note that, while graph
summarization is compression, the focus is not on finding the smallest possible representation.
Instead a graph summary is an effort to simplify the complex structure such that relations between
some represented information is revealed.

Based on the separator technique mentioned by Deo and Litow~\cite{Deo98_structuralGraphCompression},
Blandford et
al.~\cite{Blandford03_seperableGraphRepresentations,Blandford2006_compactStructuresFastQuery} design
a compressed data structure. The idea is to find graph components that can be disconnected from the
rest of the graph by only removing a small number of edges. According
to~\cite{Claude10_compactWebGraphRep} the improvements this achieves over a plain representation of
the adjacency list mainly rely on smart caching and it is therefore unclear if it would be useful on
large web graphs. They do evaluate their technique on several different graphs, including 3D meshes,
circuits, street maps, routing maps and two version of a Google web graph (once with reversed edges).
The latter is a graph with 916 thousand nodes and 5.1 million edges, of which they report compression of
9.9~bpe and 7.86~bpe with reversed edges. As this method predates every other implementation
discussed here, a comparison with other methods is not provided.
%
%
\subsection{RDF Compression}
RDF graphs are a fairly recent graph data format. It is often used to represent semantic
information. The idea is to provide the semantic information in a format that can be interpreted by
machines, ultimately aiding knowledge discovery. RDF in its purest form is a set of triples of
subject, predicate and object.

All three -- subject, predicate, and object -- allow for strings as values. Explicitly storing the
triples is therefore very memory intensive. Thus it is common practice to replace these concrete
values by some short encoding and keep a dictionary mapping the code to the concrete value. This
leads to two common approaches to compress RDF files: one is to compress the dictionary, the other
to compress the underlying graph structure.

Fern\'andez et al.~\cite{DBLP:conf/www/FernandezGM10} compare four basic approaches to compressing
RDF data on three data sets: Billion Triples (541.5~MB), Uniprot (239.4~MB) and US Census data
(148.2~MB). The compression ratios compare the original uncompressed file size
with the compressed one. Lower percentages mean better compression. The approaches are:
\begin{itemize}
    \item Directly compressing the original file using standard methods (gzip, bzip2 and ppmdi-6). Of
        those, ppmdi-6 gave the best results with compression ratios of 3.1\% to 7\%.
    \item Representing the RDF-file as an adjacency list can be done in three ways: each the
        subject, the predicate or the object can be used as the head of the list. A triple $(s,p,o)$
        can therefore be represented by the adjacency lists $s \rightarrow (p,o)$, $p \rightarrow
        (s,o)$, and $o \rightarrow (s,p)$. They evaluated all three representations and again
        compressed the resulting adjacency list using ppmdi-6. Using the subject as head worked
        best, achieving compression ratios of 2.2\% to 6.6\% 
    \item The final two methods implement the same approach using two different techniques. In both
        cases a dictionary is built from the triples as explained above. The triples are then again
        represented as an adjacency list which is Huffman-encoded. They evaluate two different
        methods of representing the dictionary: (1) delta-codes and (2) a succinct tree structure,
        where the tree is represented as a string of balanced parentheses. They both are then again
        compressed using standard compression methods. The delta-codes work best achieving 3.8\% to
        7\% compression ratios.
\end{itemize}
Mart\'inez-Prieto et al.~\cite{DBLP:conf/sac/Martinez-PrietoFC12} focus on compressing the RDF
dictionary. They develop a compressor which can be optimized for either compression or for querying.
Among others they also evaluate on a Uniprot dataset (9.11~GB file size, however using an
uncompressed dictionary already reduces this to 1.21~GB) and achieve a compression ratio of $27.61\%$
on it with the compressor optimized for compression. In this case compression ratio is calculated by
dividing the compressed dictionary size by the original dictionary size. Optimizing the compressor
for querying leads to compression that takes approximately twice the space but has two to three
times faster access times. Another approach on dictionary compression is by Urbani et
al.~\cite{DBLP:journals/concurrency/UrbaniMDSB13}. They implement an parallel approach using the
MapReduce method. On a Uniprot dataset (230.9~GB) they achieve a compression ratio of about 12\%.
Note that this number is not directly comparable to the previous approach as it represents a
comparison of the compressed data structure to the raw file, while the compression ratio in the
previous method was comparing the compressed dictionary with an uncompressed dictionary that already
uses much less space then the raw file. Another parallel approach is developed by Cheng et
al.~\cite{DBLP:journals/corr/ChengMKWT14}. They present a scalable solution for compressing RDF
data. Instead of using a single dictionary they encode triples in parallel with multiple
dictionaries achieving a compression ratio of about 22\% on a Uniprot dataset (797~GB). 

Contrary to the previous works, Jiang et al.~\cite{xia13} focus on compressing the underlying graph
structure. They propose two compression schemes (``equivalent'' and ``dependent'' compression) to
compress annotated graphs called \emph{Typed Object Graphs}. These are RDF-graphs with type
information attached to each node. Thus the edges are quintuples in this model instead of the usual
RDF-triples, as subject and object also get a type attached. The equivalent compression aims to
combine nodes that have the same type and the same family (i.e. the same neighbors and the same
relations to these) into one node. Dependent compression searches for nodes that
have only one neighbor and combines the information stored in this node with the information stored
in the node of their only neighbor. They define two metrics for compression: $\text{CR}_q$ is the
compression ratio of the quintuple compression. This is calculated by dividing the number of
quintuples in the original graph by the number of quintuples in the compressed graph. Analogously
$\text{CR}_i$ is the node compression. They evaluate both compression approaches on 5 datasets with
100 thousand to 6 million edges. The dependent compression consistently yields better results,
achieving quintuple compression ratios of 0.83 to 0.06 and node compression ratios of 0.95 to 0.18.
As a specific example they achieve 40\% quintuple compression and 20\% node compression on the
Jamendo dataset, which has 373 thousand quintuples and 281 thousand nodes.

Similarly, Pan et al.~\cite{DBLP:conf/jist/PanGRWWZ14} also focus on compressing the graph structure
by mining for redundant graph patterns. They describe two methods of removing redundancies:
\begin{itemize}
    \item By logical compression they mean the search for patterns that appear often and can be
        replaced by a generalized triple. For example, if the pattern
        \[<?x,a,\text{foaf: Person}>, <?x, a, \text{dbp: Person}>\]
        appears often, a type $T$ could be introduced, along with a rule to expand $T$ to
        $\text{foaf: Person}$ and $\text{dbp: Person}$. Then the single triple $<?x, a, T>$ would
        represent the above pattern. 
    \item RDF files are a textual representation of the graph structure. As there are many ways of
        serializing such a graph, some of these serializations are more concise than others. They
        therefore describe a way of using graph patterns to group triples in such a way that they
        can then be serialized more concisely, resulting in shorter files.
\end{itemize}
The two methods are joined by a pattern mining phase at the beginning to find the patterns which are
used to achieve the compression. Then all three steps (pattern mining, logical compression,
serialization) can be iterated to achieve compression. The
results they give are achieved by only one iteration. They evaluate their method on four data sets
with 431 thousand to 94 million RDF triples, achieving compression rates (original number of triples/compressed
number of triples) of 1.7 to 2.8. On the Jamendo dataset, which has 1 million triples, they achieve a
compression ratio of 1.72. Again, they seem to use a different version of this (and other) dataset
than the previous work, as the number of triples is not the same.
%

A different problem is considered by Fern\'andez et al.~\cite{DBLP:conf/esws/GarciaASFC14} in the
compression of streaming data. They propose a method based on Zlib to achieve lossless compression
of RDF streams. The idea is to use the structural similarities among items in an RDF stream by
combining a differential item encoding with the general purpose compression of Zlib. An evaluation
of their method achieves 10 to 30\% smaller streams than would be achieved by directly storing the
stream in a Zlib-compressed file.
\section{Lossy Compression and Reachability}
Another approach to compression is based on the assumption that only certain queries are of
interest. The idea is, to remove redundant information irreversibly, leaving just enough information
to answer the queries in question. 

Fan et al.~\cite{DBLP:conf/sigmod/FanLWW12} propose such query-preserving lossy compression for
reachability and graph pattern queries. They also consider algorithms that preserve the compressed
representation in the case of incremental changes to the graph. In their experimental analysis they
show that this method can reduce the size of a graph for reachability queries by $95\%$ and for
graph pattern queries (a variant of subgraph matching queries, matching graph patterns in terms of
bounded simulation) by $56\%$. Zhang et al.~\cite{DBLP:conf/semweb/ZhangDYZ14} apply the concept to
RDF data. They are able to achieve compression ratios of 5\% to 30\%, considering subgraph matching
as queries.

A wide range of algorithms is concerned with the answer to reachability queries on large graphs.
Commonly the procedure is to build an index structure, which is sufficient to answer reachability
queries, but can not be used to rebuild the original graph. A frequently used preprocessing step is
to construct a DAG in which every strongly connected component $c$ of the original graph is replaced
by a node, as every node in $c$ can reach any other node in $c$. Therefore reachability on DAGs is
equivalent to reachability on arbitrary graphs and it is often assumed that the input structures are
DAGs. Yu and Cheng~\cite{DBLP:series/ads/YuC10} provide a survey on reachability queries in large
graphs, we will further only mention a few newer results and otherwise point to this survey. Yildirim et
al.~\cite{2012-grail-jvldb} present \emph{GRAIL}, an indexing structure which works well on very
large graphs, while not being the best approach for smaller graphs. They use an interval labeling,
which effectively solves reachability for trees. The idea is that the nodes are labeled by
intervals, such that the question if $v$ can reach $u$ can be answered by testing for interval
containment, i.e. reachability is given if and only if the interval of $u$ is contained in the
interval of $v$. This method is often extended to DAGs by labeling a subtree of the DAG and then
supplementing the index structure to cover the non-tree edges. GRAIL on the other hand is an effort
to directly label the DAG instead of a subtree, which has the opposite problem: it can introduce
false positives, which have to be handled separately. One way they pursue this problem is by
introducing multiple intervals on every node instead of only one. These are created by taking random
walks through the graph, and drastically reduce the number of false positives. For the remaining
exceptions they propose the use of exception lists, but conclude that they do not scale well to
large graphs. Instead they use a smart DFS based recursive pruning approach. If an exception is
possible, they check if the neighborhood can reach the target and return false if this is false for
the entire neighborhood. The authors mention that more sophisticated approaches work better on small
graphs, their approach scales better to graphs with millions of nodes and edges. They experimentally
show that both the index construction as well as the query time are much faster (up to a factor of
40 for construction and over 200 for querying) than the closest competitors on large graphs, however
their index tends to be larger by a factor of 2 to 3 as well. The same authors extend the system to
one for dynamic graphs~\cite{2013-dagger}. Some of the techniques used in GRAIL (particularly
pruning of labels) are also extended into a data structure called \emph{PReaCH} by Merz and
Sanders~\cite{DBLP:conf/esa/MerzS14}, who combine it with the idea of \emph{contraction hierarchies}
to achieve a new data structure for the answer of reachability queries. Contraction hierarchies are
a technique originally used to speed up shortest path queries. Reachability contraction hierarchies
repeatedly remove nodes of in- or out-degree $0$. This makes it possible to mark the incident edges
for forward (or backward when the out-degree is 0) exploration. If no bidirectional exploration is
done, this achieves that only a subset of the edges have to be considered. Their experiments show
better results than GRAIL (factor 3 to 13 faster index construction and factor 2 to 24 for
querying), again particularly on large graphs. Both GRAIL and PReaCH also guarantee linear
preprocessing time and space. PReaCH seems to be the current state of the art for reachability
queries on large directed graphs.

A generalization of reachability queries is studied by Zou et
al.~\cite{DBLP:journals/is/ZouXY0XZ14}. They study \emph{label-constraint reachability queries},
where a set of labels is given and only edges with a label from this set may be used to determine
reachability. They propose a transformation of an edge-labeled directed graph into an augmented DAG
replacing the maximal strongly connected components as bipartite graphs. Then a Dijkstra-like
algorithm can used to answer reachability by assigning weights to the labels. As the partition-based
method for transforming the graph is $\NP$-hard, they also propose a sampling based approximation.
Their evaluation is done on 2 rather small real world graphs (at most 5 thousand nodes and 66 edge
labels) and synthetic graphs using the Erd\H{o}s-R\'{e}nyi model.
Wei et al.~\cite{DBLP:journals/pvldb/WeiYLJ14} propose the use of a different node labeling approach
that can be used to test reachability by checking for set containment. As this is time consuming for
large sets, they use a randomized approach to check set containment. If this does not lead to an
answer that is correct with 100\% probability, a DFS search is started to answer the query. Their
approach can not give the same worst-case guarantees that GRAIL provides, but they show it to work
faster in practice at both index construction and query answering time.
\section{Other related Works}
Almost every implemented lossless compression method we discuss above considers querying the compressed
graph in some way and is therefore suitable as an in-memory representation of the graph. The method
by Kang and Faloutsos~\cite{Kang11_HubsSpokesCompression} in contrast is does not consider access to
the graph. The method recursively removes the nodes of highest degree and uses the resulting
structure to reorder the adjacency matrix aiming for a representation that groups all the 0-values
into the same regions. Due to their recursive structure, this works particularly well on power-law
graphs. The final compression then relies on standard gzip-compression (which cannot be queried).
They achieve results of about 7 to 16~bpe. One advantage of this method is its suitability for for
block based matrix multiplication, which can be used to execute graph mining algorithms over
MapReduce.

Nourbakhsh et al.~\cite{nou14} also propose a compression method but do not consider access to the
compressed structure. They formulate graph compression as a matrix factorization problem and
prove that solving a continuous relaxation of this discrete optimization problem, yields an optimal
result for the original problem. They then present a novel algorithm to approximate this solution
and use it to compress the adjacency matrix representation of undirected, edge-weighted graphs with
$n$ nodes (thus a matrix of size $n^2$) to a representation of size $(n+k^2)$ for some $k < n$.
The method is not necessarily lossless, as not every matrix can be represented by a smaller
factorization.

Different approaches are needed to deal with large graphs in \emph{external memory}. Working on graphs with
external memory is quite difficult, as there is no obvious way of partitioning them into the block
structure of external memory. Kyrola et al.~\cite{DBLP:conf/osdi/KyrolaBG12} develop GraphChi,
a library to do large-scale graph computation on external memory efficiently. For example computing
100 iterations of the PageRank algorithm on a graph with 3.8 billion edges takes 581 minutes on a
machine with an SSD, which is only about 4 times slower than the same computation on a cluster of 30
machines. It is essentially an external memory implementation of
GraphLab~\cite{DBLP:conf/uai/LowGKBGH10}. Kyrola and Guestrin~\cite{DBLP:journals/corr/KyrolaG14}
improve this through a new data structure -- Parallel Adjacency Lists -- which is based
on the graph storage model of GraphChi.

Algorithms optimized on I/O-efficiency also exist. Traversing large graphs
breadth-first in external memory is considered by Beckmann et al.~\cite{DBLP:conf/esa/BeckmannMV13}
and they present an I/O-efficient implementation. Their algorithm uses a clustering
method grouping nodes into a hierarchy. Furthermore they consider dynamic data and are
able to update the BFS levels without recomputing the whole traversal from scratch. Closely related
to reachability but generally more difficult is the question of finding a shortest path between two
nodes in a weighted Using external memory is difficult for several reasons (cf.
Meyer~\cite{DBLP:conf/birthday/Meyer09}), but Meyer and Zeh achieve some results on I/O-efficient
shortest path algorithms, first for undirected graphs with bounded edge
lengths~\cite{DBLP:journals/talg/MeyerZ12}, which they then extend to graphs with unbounded edge
lengths~\cite{DBLP:conf/esa/MeyerZ06}. These algorithms are based on I/O-efficient BFS traversals by
Meyer and Mehlhorn~\cite{DBLP:conf/esa/MehlhornM02}.
\bibliographystyle{alpha}
\bibliography{bib.bib}
\end{document}